\documentclass[conference]{IEEEtran}

\usepackage{graphicx,psfrag,epsfig,epsf,latexsym,hhline,amsmath,amssymb,multirow}
\usepackage{pst-plot}
\usepackage{pstricks-add}
\usepackage{pifont}
\usepackage{graphicx}
\usepackage{amsthm}
\usepackage{algorithm}
\usepackage{algorithmic}
\usepackage[noadjust]{cite}
\usepackage{array}
\usepackage{blindtext}
\usepackage{etoolbox}
\usepackage{comment}
\usepackage{epstopdf}
\newcolumntype{M}[1]{>{\centering\arraybackslash}m{#1}}
\newcolumntype{P}[1]{>{\centering\arraybackslash}p{#1}}

\newcommand{\iid}{i.\@i.\@d.\ }

\theoremstyle{definition}\newtheorem{lemma}{Lemma}
\theoremstyle{definition}
\theoremstyle{definition}
\theoremstyle{definition}

\newtheorem{remark}[lemma]{Remark}

\begin{document}
\title{Density Evolution Analysis of Partially Information Coupled Turbo Codes on the Erasure Channel}
\author{Min Qiu, Xiaowei Wu, Yixuan Xie, and Jinhong Yuan \\
School of Electrical Engineering and Telecommunications\\
The University of New South Wales, Sydney, Australia\\
E-mail: \{min.qiu, xiaowei.wu, yixuan.xie, j.yuan\}@unsw.edu.au

%
%
}

\maketitle

\begin{abstract}
In this paper, we investigate the performance of a class of spatially coupled codes, namely partially information coupled turbo codes (PIC-TCs) over the binary erasure channel (BEC). This class of codes enjoy flexible code rate adjustment by varying the coupling ratio. Moreover, the coupling method can be directly applied to any component codes without changing the encoding and decoding architectures of the underlying component codes. However, the theoretical performance of PIC-TCs has not been fully investigated. For this work, we consider the codes that have coupling memory $m$ and study the corresponding graph model. We then derive the exact density evolution equations for these code ensembles with any given coupling ratio and coupling memory $m$ to precisely compute their belief propagation decoding thresholds for the BEC. Our simulation results verify the correctness of our theoretical analysis and also show better error performance over uncoupled turbo codes with a variety of code rates on the BEC.
\end{abstract}

\section{Introduction}
Spatially coupled codes, originally introduced in \cite{782171} as convolutional low-density parity-check (LDPC) codes, have now been recognized as promising candidates for a range of applications such as optical communication and data storage systems \cite{7553579}. These codes are constructed from a sequence of component codes coupled in a certain structure. One advantage of these codes is that they can be efficiently decoded by a window decoder in a component-wise manner, enabling a continuous streaming fashion with lower decoding delay compared to the conventional long block codes. Moreover, spatially coupled codes have been shown to achieve better decoding thresholds and lower error floors than their uncoupled counterparts. As a result, they have attracted considerable interest in both academia and industry.

Several classes of spatially coupled codes have been proposed in the literature. Some well-known examples such as spatially coupled LDPC (SC-LDPC) codes \cite{5571910,5695130,7553579}, spatially coupled turbo-like codes (SC-TCs) \cite{8002601} whose component codes includes parallel concatenated convolutional codes (PCCCs) \cite{397441} and serial concatenated convolutional codes (SCCCs) \cite{669119}, braided convolutional codes (BCCs) \cite{5361461} and staircase codes \cite{Smith12,8425763}, have all reported the close-to-capacity performance. Most importantly, it has been proved in \cite{5695130} and \cite{8002601} that the belief-propagation (BP) decoding thresholds of the SC-LDPC and SC-TC ensembles can converge to the maximum-a-posteriori (MAP) decoding thresholds of their respective underlying uncoupled ensembles. Such a phenomenon is known as threshold saturation \cite{5695130}, meaning that suboptimal BP decoding is sufficient to universally achieve the capacity of general binary-input memoryless output-symmetric channels \cite{6589171}. An intuitive explanation for this can be that the portions of the code that have already been successfully decoded can help their neighboring parts by propagating reliable information to them. Although both SC-LDPC codes and SC-TCs have comparable error performance, SC-TCs have much lower computational complexity in encoding and thus may be more favorable in practical systems.

For this work, we study a particular class of spatially coupled codes whose component codes are PCCCs. Very recently, partially information coupled turbo codes (PIC-TCs) have been proposed in \cite{8368318} where a portion of the information bits of consecutive component turbo code codewords are coupled. Different from the spatially coupled PCCCs (SC-PCCCs) in \cite{8002601}, the overall code rate can be changed flexibly by varying the coupling ratio. Simulation results therein show that the PIC-TCs constructed from LTE turbo codes can have a SNR gain up to 0.73 dB over the uncoupled LTE turbo codes. However, only coupling memory $m=1$ was considered in the design. Furthermore, the theoretical performance of the codes has not been fully understood and investigated. Although an extrinsic information transfer (EXIT) chart analysis was presented, the estimated threshold is an upper bound of the true decoding threshold because it was assumed that the pre- and the post-coupled information bits are perfectly known to the decoder when deriving the EXIT functions. As a result, the estimation of the decoding threshold is not accurate and the simulation results in Fig. 8 of \cite{8368318} even shows a large gap between the predicted and the actual decoding threshold.

In this paper, we focus on studying the performance of PIC-TCs over the BEC. First, we introduce a general construction where the PIC-TCs have coupling memory $m\geq 1$. We then look into the corresponding graph model of the code ensembles which has not been discussed in \cite{8368318}. Based on the graph model, we derive the exact density evolution (DE) equations for PIC-TC ensembles with any given coupling ratio and the coupling memory $m$ and compute their BP thresholds. Our analysis shows that the codes are capable of approaching the BEC capacity universally for a wide range of coupling ratios and code rates. Simulation results confirm our theoretical analysis and also show lower bit error rate of our PIC-TCs over the benchmark schemes.

\begin{figure}[ht!]
	\centering
\includegraphics[width=3.23in,clip,keepaspectratio]{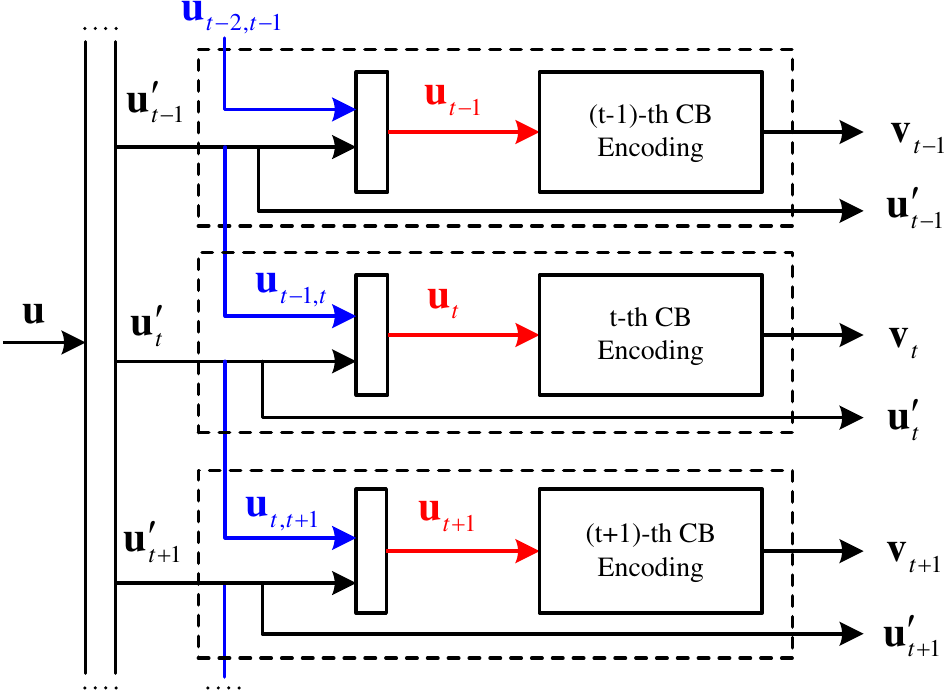}
\vspace{-4mm}
\caption{Block diagram of a PIC-TC.}
\vspace{-4mm}
\label{fig:1}
\end{figure}

\section{Partially information coupled Turbo Codes}
In this section, we introduce the encoding and decoding of PIC-TCs. We first describe the architecture of PIC-TCs with coupling memory $m = 1$. After that, we will show encoding and decoding algorithms with a given coupling memory $m$.

\vspace{-2mm}
\subsection{Encoding of PIC-TCs}
The block diagram of the PIC-TC encoder with $m = 1$ is depicted in Fig. \ref{fig:1}.

We consider that the underlying component code is a binary linear code of length $N$ and cardinality $2^{K}$. As shown in Fig. \ref{fig:1}, an information sequence $\mathbf{u}$ is divided into $L$ vectors $\mathbf{u}'_1,\ldots,\mathbf{u}'_L$, which will be encoded into $L$ code blocks (CBs), i.e., turbo codes, respectively. Each vector can be decomposed as $\mathbf{u}'_t = [\mathbf{u}_{t,t}, \mathbf{u}_{t,t+1}]$ for the time instance $t = 1,\ldots,L$, where $\mathbf{u}_{t,t}$ represents the uncoupled information sequence and $\mathbf{u}_{t,t+1}$ represent the coupled information sequence, i.e., the information shared between the $t$-th CB and the $(t+1)$-th CB. In this way, all the coupled information sequences are encoded \emph{twice} and are protected by two component turbo code codewords. Then, the input of the $t$-th CB encoder is a length $K$ vector $\mathbf{u}_t = [\mathbf{u}_{t-1,t},\mathbf{u}'_t]$. The parity bits of the $t$-th CB are represented by a length $N-K$ vector $\mathbf{v}_t$. In this work, we consider the underlying component code to be a rate $\frac{K}{N} = \frac{1}{3}$ turbo code although the choices of component code can be any systematic codes, e.g., polar codes \cite{8470926}. We denote by $D = |\mathbf{u}_{t,t+1}|$ the length of the coupling information sequence. The coupling ratio $\lambda = \frac{D}{K} \in [0,\frac{1}{2}]$ is an important parameter that determines the overall code rate and the decoding threshold, which will be discussed later.

\begin{table}[ht!]
\begin{algorithm}[H]
\normalsize
\caption{PIC-TC Encoding}\label{alg:en}
\begin{algorithmic}[1]
\STATE Divide $\mathbf{u}$ into $\mathbf{u}'_1,\ldots,\mathbf{u}'_L$
\FOR {$t = 1 \to L$}
\STATE Decompose $\mathbf{u}'_t$ into $\mathbf{u}_{t,t},\mathbf{u}_{t,t+1},\ldots,\mathbf{u}_{t,t+m}$
\STATE Construct the information sequence for the $t$-th CB: \\
$\mathbf{u}_t = [\mathbf{u}_{t-m,t},\ldots,\mathbf{u}_{t-1,t},\mathbf{u}'_t] $\\
$\hspace{4.5mm}= [\mathbf{u}_{t-m,t},\ldots,\mathbf{u}_{t-1,t},\mathbf{u}_{t,t},\mathbf{u}_{t,t+1},\ldots,\mathbf{u}_{t,t+m}]$
\FOR {$j = 1 \to m$}
\IF {$t>L-j$}
\STATE $\mathbf{u}_{t,t+j} =\mathbf{0}$
\ELSIF {$t<j$}
\STATE $\mathbf{u}_{t-j,t} = \mathbf{0}$
\ENDIF
\ENDFOR
\STATE Perform turbo encoding on $\mathbf{u}_t$ to obtain $\mathbf{v}_t$
\STATE Construct the codeword of the $t$-th CB: $[\mathbf{u}'_t,\mathbf{v}_t]$
\ENDFOR
\end{algorithmic}
\end{algorithm}
\vspace{-5mm}
\end{table}

For coupling memory $m \geq 1$, the coupled information sequence is divided into $m$ sequences of length $D/m$ $\mathbf{u}_{t,t+1},\ldots,\mathbf{u}_{t,t+m}$, which are fed into the $(t+1)$-th CB$,\ldots,$$(t+m)$-th CB, respectively. The encoding procedure is described in Algorithm \ref{alg:en}. Here, Steps 5-10 of Algorithm \ref{alg:en} is the zero padding process which is similar to the case of SC-PCCCs in \cite{8002601}. Given the parameters of the component code $(K,N)$, the number of CBs $L$ and the coupling ratio $\lambda$, the overall code rate is
\begin{align}
R_{\text{PIC}} &\overset{(a)}= \frac{L(K-\lambda K)-\lambda K}{L(N-\lambda K)-(m\lambda K - \frac{\lambda K}{m}(2^{m-1}-1))} \label{eq:PICrate} \\
& \overset{L \rightarrow \infty}= \frac{R-\lambda R}{1-\lambda R}, \label{eq:PICrate2}
\end{align}
where $(a)$ is based on the code rate given in \cite[Eq. (9)]{8368318} by considering coupling memory $m$ and subtracting the number of bits for zero padding. It is obvious that the rate loss due to the zero padding will become negligible when $L$ is large.

\begin{remark}
It can be seen from \eqref{eq:PICrate2} that the code rate of a PIC-TC can be lowered by increasing $\lambda$. This is because the coupled information bits are encoded by two turbo encoders (i.e., four convolutional code encoders), which is different from the SC-PCCCs in \cite{8002601} where the coupled bits are encoded by two convolutional code encoders. That is, the coupling of our PIC-TCs is in the turbo code level while the coupling for the SC-PCCCs in \cite{8002601} is in the convolutional code level. Therefore, our coupling method can be directly applied to any component code \emph{without changing} its encoder and decoder. Consequently, the performance analysis of our PIC-TCs in Section \ref{sec:per_ans} is also different from that of SC-PCCCs in \cite{8002601}. It will be shown later that our PIC-TCs can still maintain the close-to-capacity performance for lowered code rates without completely redesigning the underlying component turbo codes. 
\end{remark}

\subsection{Decoding of PIC-TCs}
The decoding of PIC-TCs is accomplished by a feed-forward and feed-back (FF-FB) decoding \cite{8368318} in an iterative manner. In short, it employs a serial scheduling by decoding the first CB to the last CB serially and then starts from decoding the last CB to the first CB if necessary. Compared to the conventional window decoding, it suffers from larger decoding delay but with improved error performance.

\begin{table}[ht!]
\begin{algorithm}[H]
\normalsize
\caption{PIC-TC Decoding}\label{alg:de}
\begin{algorithmic}[1]
\FOR {$i = 1 \to I_{\max}$}
\FOR {$t = 1 \to L$}
\FOR {$j = 2 \to m$}
\IF {$t>L-j$}
\STATE $\mathbf{L}^{(\text{out})}_{t,t+j} =\infty$
\ELSIF {$t<j$}
\STATE $\mathbf{L}^{(\text{in})}_{t-j,t} = \infty$
\ELSE
\STATE $\mathbf{L}^{(\text{in})}_{t-j,t} = \mathbf{L}^{(\text{out})}_{t-1-j,t-1}$
\ENDIF
\ENDFOR
\STATE Compute the LLRs for $\mathbf{u}_t$: \\
$[\mathbf{L}^{(\text{in})}_{t-m,t},\ldots,\mathbf{L}^{(\text{in})}_{t-1,t},\mathbf{L}^{(\text{in})}_{t,t},\mathbf{L}^{(\text{in})}_{t,t+1},\ldots,\mathbf{L}^{(\text{in})}_{t,t+m}]$
\STATE Input all the LLRs into the turbo decoder and obtain: $[\mathbf{L}^{(\text{out})}_{t-m,t},\ldots,\mathbf{L}^{(\text{out})}_{t-1,t},\mathbf{L}^{(\text{out})}_{t,t},\mathbf{L}^{(\text{out})}_{t,t+1},\ldots,\mathbf{L}^{(\text{out})}_{t,t+m}]$
\ENDFOR
\STATE Perform Steps 3-13 for $t = L \to 1$.
\ENDFOR
\end{algorithmic}
\end{algorithm}
\vspace{-5mm}
\end{table}

The decoding for each CB is realized by a standard turbo decoder with the BCJR decoder \cite{1055186} as the constituent decoder. We denote by $\mathbf{L}^{(\text{in})}_{t,j}$ and $\mathbf{L}^{(\text{out})}_{t,j}$ the turbo decoder input and output log-likelihood ratio (LLR) sequence for $\mathbf{u}_{t,j}$, respectively. The maximum number of iterations is denoted by $I_{\max}$. The FF-FB decoding steps for PIC-TCs with $m$ is given in Algorithm \ref{alg:de}. Under the BEC, the bits used for zero padding will have the LLR values of $\infty$ at the input of the decoder.

\section{Performance Analysis of PIC-TCs}\label{sec:per_ans}
In this section, we analyze the decoding performance of PIC-TCs by using density evolution. We first look into the corresponding graph model of the PIC-TC ensembles. The exact DE equations are then derived based on the graph model.

\subsection{Graph Model Representation}
\begin{figure}[ht!]
	\centering
\includegraphics[width=3.43in,clip,keepaspectratio]{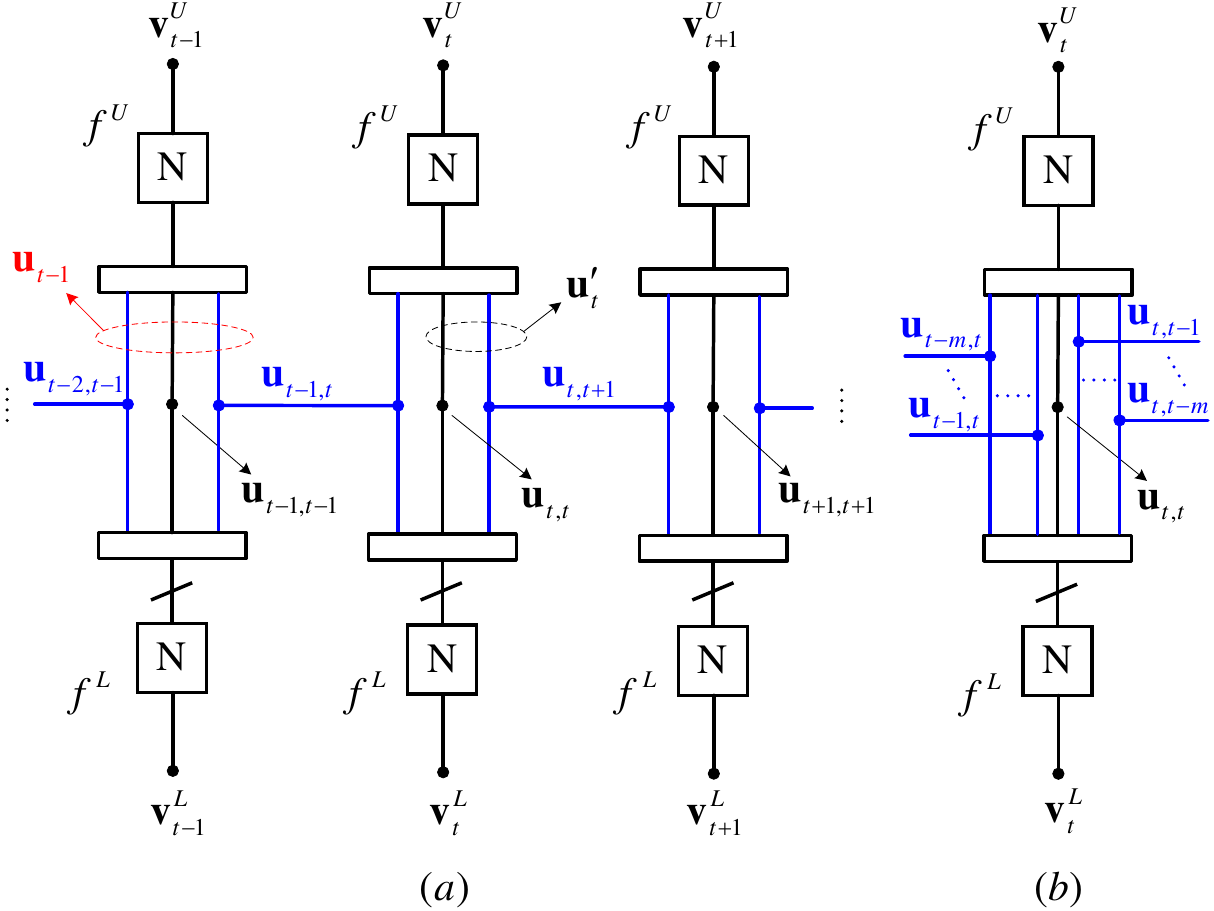}
\vspace{-3mm}
\caption{Compact graph representation of (a) PIC-TCs with coupling memory $m = 1$ from time instant $t-1$ to $t+1$, and (b) PIC-TCs of coupling memory $m>1$ for time instant $t$.}
\label{fig:2}
\vspace{-5mm}
\end{figure}

We start with the case of PIC-TCs with $m=1$. We note that the turbo code ensembles can be represented by a compact graph \cite[Sec. III]{8002601}, which simplifies the factor graph representation. The main idea is that each of the sequences of information bits and parity bits in the factor graph is represented by a single variable node and the trellises are represented by factor nodes. Based on this idea, we use the compact graph representations on the PIC-TC ensembles, which are depicted in Fig. \ref{fig:2}.

Since the component code is a turbo code built from two rate-$\frac{1}{2}$ recursive systematic convolutional codes, the corresponding compact graph of each turbo code has an upper decoder and a lower decoder. Two factor nodes $f^U$ and $f^L$ represent the upper and lower convolutional code decoders, respectively while the ``$N$'' inside the node means the decoder takes a length $N$ sequence as its input. The parity sequences input to upper and lower decoder at time $t$ are denoted by $\mathbf{v}^U_t$ and $\mathbf{v}^L_t$, respectively. The effect of interleaving on the information sequence of the turbo code is represented by a slash on the edge between the information nodes and the lower factor node.

As shown in Fig. \ref{fig:2}(a), for $m=1$ and at time $t$, we use three information nodes to represent $\mathbf{u}_t$\footnote{To ease the notation, we refer to an information node representing a sequence as the sequence itself.}, i.e., by treating the coupled and uncoupled information sequences separately. The coupled information sequence $\mathbf{u}_{t-1,t}$ is part of the input of the decoders at time $t-1$ and time $t$. Hence, the coupled information nodes $\mathbf{u}_{t-1,t}$ are in both of the compact graphs at time $t-1$ and time $t$, respectively and they are connected by a horizontal line as shown in the figure. In other words, since the coupled information sequences $\mathbf{u}_{t-1,t}$ are encoded by four convolutional encoders at time $t-1$ and time $t$, the extrinsic information is passed between \emph{the upper and lower} factor nodes at time $t-1$ and time $t$ \emph{via} node $\mathbf{u}_{t-1,t}$. Similarly, the coupled information nodes $\mathbf{u}_{t,t+1}$ connect the compact graphs at time $t$ and time $t+1$, respectively. However, the uncoupled information node $\mathbf{u}_{t,t}$ only connects the upper and lower factor nodes at time $t$ as the uncoupled information bits are only encoded by the two constituent convolutional encoders at time $t$. The extrinsic information from $\mathbf{u}_t$ at time $t$ will be propagated to the upper or the lower factor nodes at time instances $t-1,t,t+1$ simultaneously.

Fig. \ref{fig:2}(b) shows the compact graph representation of the PIC-TC ensemble with coupling memory $m$. Here, $\mathbf{u}_t$ is represented by $2m+1$ information nodes in the graph where the leftmost $m$ information nodes connect to $m$ compact graphs from time $t-m$ to time $t-1$ and similarly for the rightmost $m$ information nodes.

\subsection{Density Evolution}\label{sec:DE}
In this subsection, we derive the exact DE equations for the PIC-TC ensembles and analyze the decoding threshold. We denote by $\epsilon$ the channel erasure probability.

For transmission over the BEC, the asymptotic behavior of the PIC-TCs can be analyzed exactly by tracking the evolution of the erasure probability over decoding iterations. 

1) We first look at the PIC-TCs with $m=1$. For the compact graph at time $t$, we let $p^{(i)}_{U,t}$ and $q^{(i)}_{U,t}$ represent the average extrinsic erasure probability from $f^U$ to $\mathbf{u}_t$ and $\mathbf{v}^U_t$, respectively, after $i$ iterations. Similarly, we let $p^{(i)}_{L,t}$ and $q^{(i)}_{L,t}$ represent the average extrinsic erasure probability from $f^L$ to $\mathbf{u}_t$ and $\mathbf{v}^L_t$, respectively, after $i$ iterations. The transfer function of the upper factor node for information bits and parity bits are denoted by $F^U_p$ and $F^U_q$, respectively. Similarly, the transfer function of the lower factor node for information bits and parity bits are denoted by $F^L_p$ and $F^L_q$, respectively. Here, the transfer function of a specific convolutional code with BCJR decoding under the BEC can be derived by using the methods proposed in \cite{1258535}. The details are omitted due to space limit. 

We denote by $\bar{p}^{(i)}_{L,t}$ the average erasure probability of all the information nodes, i.e., $\mathbf{u}_t$, input to the factor node $f^U$ at time $t$. Based on the graph model in Fig. \ref{fig:2}, $\bar{p}^{(i)}_{L,t}$ is the \emph{weighted sum} of the erasure probabilities of information node $\mathbf{u}_{t-1,t}$, $\mathbf{u}_{t,t}$ and $\mathbf{u}_{t,t+1}$ to node $f^U$, where the weight is determined by the coupling ratio $\lambda$. It is computed as
\begin{align}\label{eq:1}
\bar{p}^{(i)}_{L,t} 
 \hspace{-1mm}= \hspace{-1mm}\epsilon\big(\lambda\cdot p^{(i)}_{L,t-1}\cdot  p^{(i)}_{L,t} \hspace{-1mm}+  (1 \hspace{-1mm}- \hspace{-1mm}2\lambda) p^{(i)}_{L,t}\hspace{-1mm} + \hspace{-1mm} \lambda \cdot  p^{(i)}_{L,t} \cdot p^{(i)}_{L,t+1}\big).
\end{align}
where $1-2\lambda$ is the ratio of the uncoupled information sequences. Note that $p^{(i)}_{L,t-1}\cdot p^{(i)}_{L,t}$ is the extrinsic erasure probability on information node $\mathbf{u}_{t-1,t}$ that depends on both the extrinsic erasure probability from the factor node $f^L$ to $\mathbf{u}_{t-1,t}$ at time $t-1$ and the extrinsic erasure probability from node $f^L$ to the same information node $\mathbf{u}_{t-1,t}$ at time $t$. Since node $\mathbf{u}_{t-1,t}$ connects the compact graphs at time $t-1$ and time $t$ according to Fig. \ref{fig:2}(a), the erasure probabilities on the two edges between $f^L$ at time $t-1, t$ and $\mathbf{u}_{t-1,t}$ \emph{multiply} with each other. Then, the overall erasure probability on $\mathbf{u}_{t-1,t}$ is the multiplication of average extrinsic erasure probability from above and the channel erasure probability $\epsilon$. Similar to the above principle, we can also obtain the a-posteriori erasure probability on information node $\mathbf{u}_{t,t+1}$ as $\epsilon \cdot p^{(i)}_{L,t}\cdot p^{(i)}_{L,t+1}$.
The DE update at node $f^U$ at time $t$ can be written as
\begin{align}
p^{(i+1)}_{U,t} &= F^U_p(\bar{p}^{(i)}_{L,t},\bar{q}^{(i)}_{U,t}), \\
q^{(i+1)}_{U,t} &= F^U_q(\bar{p}^{(i)}_{L,t},\bar{q}^{(i)}_{U,t}),
\end{align}
where $\bar{q}^{(i)}_{U,t} = \epsilon$ is the average erasure probability of $\mathbf{v}^U_t$ to factor node $f^U$ at time $t$.

Similar to the above steps, the average erasure probability from $\mathbf{u}_t$ to node $f^L$ at time $t$ is computed as
\begin{align}\label{eq:2}
\bar{p}^{(i)}_{U,t}\hspace{-1mm} = \hspace{-1mm}\epsilon\big(\lambda\cdot p^{(i)}_{U,t-1}\cdot p^{(i)}_{U,t} \hspace{-1mm}+  (1 \hspace{-1mm}- \hspace{-1mm} 2\lambda) p^{(i)}_{U,t} \hspace{-1mm} + \hspace{-1mm} \lambda \cdot p^{(i)}_{U,t}\cdot p^{(i)}_{U,t+1}\big).
\end{align}
The DE update at node $f^L$ at time $t$ can then be written as
\begin{align}
p^{(i+1)}_{L,t} = F^L_p(\bar{p}^{(i)}_{U,t},\bar{q}^{(i)}_{L,t}), \\
q^{(i+1)}_{L,t} = F^L_q(\bar{p}^{(i)}_{U,t},\bar{q}^{(i)}_{L,t}),
\end{align}
where $\bar{q}^{(i)}_{L,t} = \epsilon$ is the erasure probability of $\mathbf{v}^L_t$ to factor node $f^L$ at time $t$.

Finally, the a-posteriori erasure probability of $\mathbf{u}_t$ at the $i$-th iteration is
\begin{align}
p_{\mathbf{u}_t}^{(i)} = \epsilon \cdot p^{(i)}_{L,t}\cdot p^{(i)}_{U,t}.
\end{align}

2) We now extend the analysis to PIC-TCs with $m>1$. Since the erasure probability of $\mathbf{u}_t$ depends on the erasure probability of the coupled information $\mathbf{u}_{t-m,t},\ldots,\mathbf{u}_{t-1,t}$ and $\mathbf{u}_{t,t+1},\ldots,\mathbf{u}_{t,t+m}$, the average erasure probability of $\mathbf{u}_t$ to $f^U$ at time $t$ from \eqref{eq:1} needs to be modified to
\begin{align}
\bar{p}^{(i)}_{L,t} 
=& \epsilon\bigg(\frac{\lambda}{m}\sum\nolimits_{j=1}^m p^{(i)}_{L,t-j}\cdot p^{(i)}_{L,t}+ (1-2\lambda) p^{(i)}_{L,t} \nonumber \\
& + \frac{\lambda}{m}\sum\nolimits_{j=1}^m p^{(i)}_{L,t}\cdot p^{(i)}_{L,t+j}\bigg),
\end{align}
where $\frac{\lambda}{m}$ is the portion of the coupled information sequence shared between two CBs in different time instances.

Similarly, the average erasure probability from $\mathbf{u}_t$ to node $f^L$ at time $t$ from \eqref{eq:2} is modified to
\begin{align}
\bar{p}^{(i)}_{U,t} =& \epsilon\bigg(\frac{\lambda}{m}\sum\nolimits_{j=1}^m p^{(i)}_{U,t-j}\cdot p^{(i)}_{U,t}+ (1-2\lambda) p^{(i)}_{U,t}   \nonumber \\
&+ \frac{\lambda}{m}\sum\nolimits_{j=1}^m p^{(i)}_{U,t}\cdot p^{(i)}_{U,t+j}\bigg).
\end{align}
The rest of the DE equations will remain the same.

We use the developed DE equations to track the evolution of the a-posteriori erasure probability on the information bits and then determined the BP decoding threshold of our PIC-TCs.

\section{Results and Discussions}\label{SIM}
In this section, we first present the theoretical decoding threshold for some PIC-TC ensembles by using the DE analysis in Section \ref{sec:DE}. After that, we show the simulation results on the error performance of our PIC-TCs.

\subsection{Density Evolution Results}
We consider that the underlying rate-$\frac{1}{3}$ turbo code is a parallel concatenation of two identical 4-state rate-$\frac{1}{2}$ convolutional codes whose generator polynomial is $(1,\frac{5}{7})$ in octal notation. As a result, the transfer functions for the upper and lower decoder in the compact graph presented in Fig. \ref{fig:2} are the same. We then compute the BP threshold of PIC-TCs built from this convolutional code for a variety of $\lambda$, code rates and with infinite decoding iterations and CB length. The BP threshold of different coupling memories (denoted by $\epsilon^{(m)}_{\text{BP}}$) and the gap to the BEC capacity (computed by $1-R_{\text{PIC}} - \epsilon^{(m)}_{\text{BP}}$) are shown in Table \ref{table3}.

\begin{table}[ht!]
  \centering
 \caption{DE thresholds FOR PIC-TCs}\label{table3}
\begin{tabular}{|c|c|c|c|c|c|}
\hline
 $R_{\text{PIC}}$  & $\lambda$   & $\epsilon^{(m=1)}_{\text{BP}}$ & $\epsilon^{(m=2)}_{\text{BP}}$ & $\epsilon^{(m=3)}_{\text{BP}}$ & Gap to capacity \\  \hline
 0.3191 & $\frac{1}{16}$ & 0.6596 & 0.6596 & 0.6596 & 0.0213 \\
   0.3043 & $\frac{1}{8}$ & 0.6756 & 0.6756 & 0.6756 & 0.0201 \\
  0.3000 & $\frac{1}{7} $ & 0.6802 & 0.6802 & 0.6802 & 0.0198 \\
   0.2941 & $\frac{1}{6} $ & 0.6862 & 0.6862 & 0.6862 & 0.0197 \\
  0.2857 & $\frac{1}{5} $ & 0.6947 & 0.6947 & 0.6947 & 0.0196 \\
   0.2727 & $\frac{1}{4} $ & 0.7075 & 0.7075 & 0.7075 & 0.0198 \\
   0.2500 & $\frac{1}{3} $ & 0.7294 & 0.7294 & 0.7294 & 0.0206 \\

   0.2381 & $\frac{3}{8} $ & 0.7406 & 0.7406 & 0.7406 & 0.0213 \\
   0.2000 & $\frac{1}{2} $ & 0.7760 & 0.7760 & 0.7760 & 0.0240 \\
  \hline
\end{tabular}
\vspace{-3mm}
\end{table}

From the table, it can be seen that the DE threshold of PIC-TCs with a range of code rates is close to the BEC capacity. In particular, the gap to the BEC capacity is about 0.02. When the code rate is very low, i.e., $R_{\text{PIC}} = 0.2$, our code can still perform very well with a gap of 0.0240. To the best of the authors' knowledge, such results have not been reported in the literature of SC-PCCCs when the code rates are lowered. It is also interesting to note that the decoding threshold of the PIC-TC with increased coupling memory remain \emph{unchanged}. This behavior is \emph{similar} to the SC-PCCCs in \cite{8368318}.

\subsection{Error Performance Results}
\begin{figure}[ht!]
	\centering
\includegraphics[width=3.42in,clip,keepaspectratio]{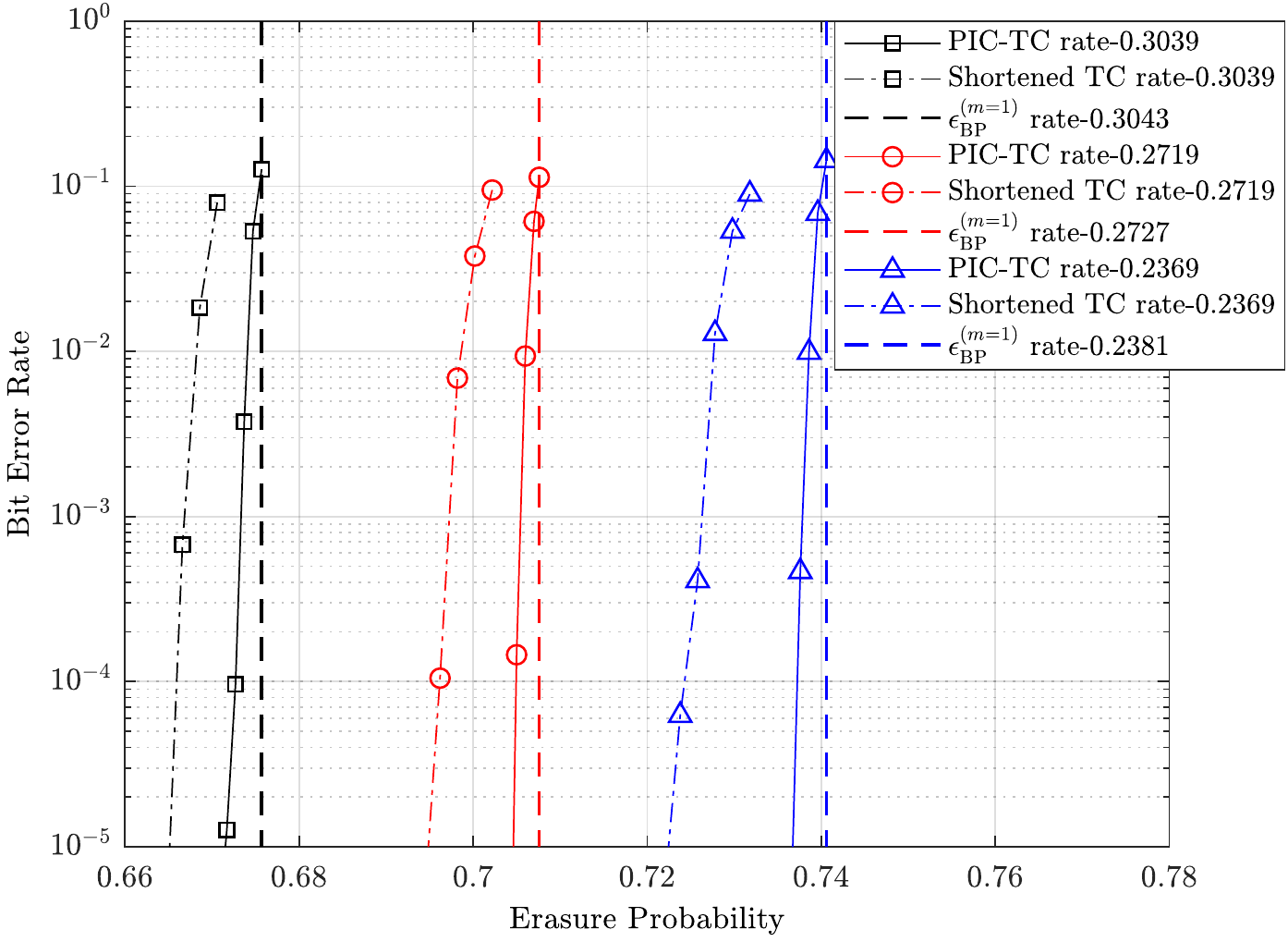}
  \vspace{-3mm}
\caption{Error performance of several PIC-TCs and shortened TCs with different code rates.}
\label{fig:3}
\end{figure}

We now present simulation results for some of the PIC-TCs with $m=1$ and finite block length. The component turbo code of our PIC-TCs is with rate $\frac{1}{3}$ and $K = 6144$. In order to minimize the code rate loss due to zero padding, we set $L = 100$. We consider three PIC-TCs with $\lambda = \frac{1}{8}, \frac{1}{4},\frac{3}{8}$, corresponding to the overall code rates of 0.3039, 0.2719, 0.2369, respectively, where the code rates are computed by \eqref{eq:PICrate}. The performance is measured in terms of bit error rate (BER) versus the erasure probability of the BEC and is shown in Fig. \ref{fig:3}. Moreover, the corresponding DE thresholds of these codes (i.e., with the same coupling ratios but slightly higher code rates due to $L \rightarrow \infty$) are included in the figure. To carry out performance comparison and since the performance of lower rate SC-PCCCs has not been reported in the literature, we plot the error performance of a number of shortened turbo codes whose mother code is with rate-$\frac{1}{3}$ and $K=100,000$, with the same code rates as that of our PIC-TCs. Although our PIC-TCs have longer block length, we note that further increasing the block lengths of the benchmark uncoupled codes only leads to negligible performance improvement.

It can be observed from Fig. \ref{fig:3} that the actual decoding performance of all the PIC-TCs is very close to the DE threshold, i.e., within 0.005 at a BER of $10^{-5}$. Most importantly, our PIC-TCs outperform the shortened turbo codes for various code rates. This implies that by simply increasing the coupling ratio to lower the code rate and without completely redesigning the code structures, our PIC-TCs can always provide better error performance over shortened turbo codes and can still approach the BEC capacity as shown in Table \ref{table3}.

\section{Concluding Remarks}\label{sec:conclude}
In this paper, we investigated the performance of PIC-TCs under the BEC, where the underlying component code is PCCCs. We considered that our PIC-TCs have coupling memory $m$ and provided the encoding and decoding algorithms for such case. We then introduced the graph model representations of our PIC-TC ensembles and derived the exact DE equations for any given coupling memory and coupling ratio. We showed that by simply varying the coupling ratio to change the overall code rate, our PIC-TCs can always approach the BEC capacity within 0.02. Simulation results verify our theoretical analysis and also demonstrated better error performance of our PIC-TCs over shortened turbo codes.

\bibliographystyle{IEEEtran}
\bibliography{MinQiu}

\end{document}